# Hemodynamic analysis of the Pulsatile Flow in Tubes of Bipolar Cross Sections

Doyeol (David) Ahn[1,2]

[1]Department of Electrical and Computer Engineering,
University of Seoul, 163 Seoulsiripdae-ro, Dongdaemun-gu, Seoul 02504, Republic of Korea
[2]Singularity Quantum Inc,
895 Dove St, Second Floor, Newport Beach, CA, 92660, USA

## ABSTRACT

Pulsatile and Poiseuille flow through compressed or defective blood vessels is a topic of fundamental importance in hemodynamics, particularly in cardiovascular research. This study examines flow dynamics within a tube with a bipolar cross-section, possibly representing the geometry of bicuspid aortic valves (BAV), aortic bifurcations, and the aortic arch—regions where non-uniform vessel shapes significantly influence hemodynamic behavior.

An analytical solution is derived for the governing equations of pulsatile and Poiseuille flow in a bipolar cross-sectional tube. The analysis focuses on the velocity field, flow rate, and wall shear stress (WSS) across different pulsation frequencies and geometric parameters, highlighting how these factors interact to shape flow characteristics.

At low frequencies, the velocity profile remains smooth, with gradual acceleration and deceleration phases. In contrast, at higher frequencies, oscillatory effects become more pronounced, and the peak volume flow, initially occurring near $\omega t = 0$ and $\omega t = \pi$, shifts toward an earlier phase in the cycle ($\omega t = 0$ to $\omega t = \pi/2$) before stabilizing at very high frequencies.

Shear stress behavior also exhibits frequency-dependent variations. At low frequencies, the fluid responds smoothly to pressure gradients, producing a shear stress distribution similar to steady flow. However, as frequency increases, inertial and unsteady effects introduce phase lags, leading to more complex shear stress patterns. These findings provide valuable insights into the interplay between vessel geometry and pulsatile forces, with implications for understanding disease progression and refining diagnostic models in cardiovascular medicine.

## I. Introduction

Pulsatile flow in a tube is a fundamental problem in biomedical engineering due to its significance in hemodynamics, particularly in understanding blood flow behavior under physiological conditions. Previous analytical studies primarily focus on tubes with circular [1,2] or elliptical cross sections [3,4], as many blood vessels, including arteries and veins, naturally exhibit these shapes. This assumption is justified in most cases, given that the vascular system is designed to maintain nearly circular lumens for optimal blood transport. However, under certain conditions—such as external tissue compression, vascular disease, or surgical interventions—the cross-sectional shape of blood vessels can deviate from a simple circular or elliptical form. These deviations can significantly alter local hemodynamics, affecting parameters such as velocity distribution, shear stress, and flow separation, which in turn influence disease progression and treatment outcomes.

In this study, we explore the use of bipolar cylindrical coordinates to model blood vessels that undergo deformation due to surrounding tissue compression and possibly bifurcation [5,6]. Unlike conventional cylindrical or elliptical models, which assume symmetric cross sections, bipolar cylindrical coordinates provide a more flexible mathematical framework for representing deformed or asymmetric vascular geometries. While patient-specific or curvilinear models derived from imaging data are ideal for clinical applications, bipolar cylindrical coordinates serve as a valuable tool for simplified analytical studies, particularly in cases where curved and branching structures are involved.

These coordinates have potential applications in studying blood flow in bicuspid aortic valves (BAV) [7-12], aortic bifurcations, and the aortic arch, where non-uniform vessel shapes impact hemodynamic patterns. Additionally, they can be used to analyze coronary artery bifurcations, where flow disturbances contribute to conditions such as atherosclerosis and plaque formation. Bipolar cylindrical models may also assist in characterizing blood flow in aneurysms and stenotic arteries. Furthermore, in cardiac flow simulations, these coordinates may help describe complex flow behavior near heart valves and within the left ventricle, particularly under conditions of altered geometry due to pathology or intervention. While bipolar cylindrical coordinates are not widely implemented in full-scale computational fluid dynamics (CFD) simulations, they hold promise for analytical hemodynamics and idealized flow studies, providing valuable theoretical insights that can complement numerical and experimental approaches.

Previous analytical solutions for pulsatile flow in circular cylindrical tubes were developed by Uchida [1], and Womersley [2], forming the foundation of blood flow analysis in large vessels. Later, Khamiri [3], Haslam, and Zamir [4] extended these studies to elliptical cross sections, accounting for cases where vessel deformation results in elongation along one axis. However, to the best of the authors 'knowledge, pulsatile flow in tubes with bipolar cylindrical cross sections has not yet been investigated in detail. Addressing this gap is essential for expanding our understanding of hemodynamics in deformed or compressed vessels, particularly in contexts where idealized circular or elliptical assumptions fail to capture the true complexity of blood flow. This study aims to fill that gap by developing a mathematical framework for pulsatile flow analysis in bipolar cylindrical coordinates, offering new insights into the role of vessel deformation in cardiovascular function and disease.

## II. Pulsatile Poiseuille Flow

When a pressure gradient $k$ acts in the axial direction of a tube of bipolar cross section, the governing equation for the pulsatile flow is [13]

$$\rho\frac{\partial u}{\partial t}+\frac{\partial P}{\partial z}=\mu\left(\frac{\partial^2 u}{\partial x^2}+\frac{\partial^2 u}{\partial y^2}\right) \tag{1}$$

where $P$ is the pressure, $u(x,y,t)$ is the axial velocity component, $\rho$ is the fluid density, $\mu$ is the fluid viscosity, and $x, y$ are rectangular coordinate along these axes. It is shown that the transverse components of the velocity are identically zero [14].

Pulsatile flow in a tube consists of an oscillatory component superimposed on a steady component, thus the velocity profile and the pressure are written by

$$\begin{gathered} u(x,y,t)=u_s(x,y)+U_\phi(x,y,t), \\ P(z,t)=P_s(z)+P_\phi(z,t). \end{gathered} \tag{2}$$

Substituting Eq. (2) into Eq. (1), we obtain

$$\{\frac{\partial P_s}{\partial z}-\mu\left(\frac{\partial^2 u_s}{\partial x^2}+\frac{\partial^2 u_s}{\partial y^2}\right)\}+\{\rho\frac{\partial u_\phi}{\partial t}+\frac{\partial P_\phi}{\partial z}-\mu\left(\frac{\partial^2 u_\phi}{\partial x^2}+\frac{\partial^2 u_\phi}{\partial y^2}\right)\}=0, \tag{3}$$

where terms have been grouped into which do not depend on time $t$ and those which depends on time. Because of that difference between them, each group must be equal to zero separately. Therefore, we have

$$\frac{\partial P_s}{\partial z}-\mu\left(\frac{\partial^2 u_s}{\partial x^2}+\frac{\partial^2 u_s}{\partial y^2}\right)=0, \tag{4a}$$

and

$$\rho\frac{\partial u_\phi}{\partial t}+\frac{\partial P_\phi}{\partial z}-\mu\left(\frac{\partial^2 u_\phi}{\partial x^2}+\frac{\partial^2 u_\phi}{\partial y^2}\right)=0. \tag{4b}$$

For the problem we are addressing, solving the equation becomes significantly more straightforward by converting to bipolar coordinates [15, 16]. This transformation tailors the coordinate system to better fit the geometric complexities often encountered in non-circular cross-section scenarios, enhancing the mathematical handling and solution accuracy of the flow dynamics.

$$x=\frac{asinh\eta}{cosh\eta-cos\xi},\qquad y=\frac{asin\xi}{cosh\eta-cos\xi}, \tag{5}$$

where $2a$ is the inter-focal distance. Here $\xi$ coordinate represents one of the two angular coordinates in bipolar coordinates. It measures the angle formed by the line connecting the point to one focus and the line perpendicular to the line connecting the two foci. Essentially, $\xi$ can be thought of as describing the angles around each focus. On the other hand, $\eta$ coordinate is the second angular coordinate and measures the logarithmic distance ratio of a point to the two foci. Fig. 1 illustrates the bipolar coordinates. The coordinate $\xi$ varies from $\xi_*$ on the upper wall of blood vessel of a bipolar cylindrical cross section to $\pi$ and from $\pi$ to $2\pi-\xi_*$ on the lower wall of a bipolar cylindrical cross section. The governing equations (4a) and (4b) rewritten in the bipolar cylindrical coordinates become

$$\frac{\partial P_s}{\partial z}-\mu\frac{(cosh\eta-cos\xi)^2}{a^2}\left(\frac{\partial^2 u_s}{\partial \eta^2}+\frac{\partial^2 u_s}{\partial \xi^2}\right)=0, \tag{6a}$$

and

$$\rho\frac{\partial u_\phi}{\partial t}+\frac{\partial P_\phi}{\partial z}-\mu\frac{(cosh\eta-cos\xi)^2}{a^2}\left(\frac{\partial^2 u_\phi}{\partial x^2}+\frac{\partial^2 u_\phi}{\partial y^2}\right)=0. \tag{6b}$$

We assume the oscillatory pressure gradient superimposed to the steady state pressure gradient of the form

$$\frac{\partial P_s}{\partial z}=k,\frac{\partial P_\phi}{\partial z}=ke^{i\omega t}, \tag{7}$$

where $\omega$ is the oscillatory frequency and $k$ is a constant which we choose to be equal to the constant pressure gradient in Poiseuille flow (6a). This choice makes it easier to compare the oscillatory flow directly with the corresponding Poiseuille flow. The exponential form of the function of time on the pressure gradient also dictates that the oscillatory velocity profile $u_\phi$ has exponential time dependence given by

$$u_\phi(\eta,\xi,t)=U_\phi(\eta,\xi)e^{i\omega t}. \tag{8}$$

From Eqs. (6a), (7) and (8), we obtain

$$\frac{(cosh\eta-cos\xi)^2}{a^2}\left(\frac{\partial^2 U_\phi}{\partial \eta^2}+\frac{\partial^2 U_\phi}{\partial \xi^2}\right)-\frac{i\omega\rho}{\mu}-\frac{k}{\mu}=0. \tag{9}$$

The governing equation for Poiseuille or steady solution is given by

$$\frac{(cosh\eta - cos\xi)^2}{a^2}\left(\frac{\partial^2 u_s}{\partial \eta^2}+\frac{\partial^2 u_s}{\partial \xi^2}\right)-\frac{k}{\mu}=0, \quad (10)$$

We assume the solution of Poiseuille flow of the form:

$$u_s(\eta,\xi)=\frac{f(\xi)}{cosh\eta - cos\xi}, \quad (11)$$

where $f(\xi)$ is an arbitrary function.

From

$$\begin{gathered}
\frac{\partial}{\partial\eta}\frac{1}{cosh\eta - cos\xi}=-\frac{sinh\eta}{(cosh\eta - cos\xi)^2},\\
\frac{\partial^2}{\partial\eta^2}\frac{1}{cosh\eta - cos\xi}=-\frac{cosh\eta}{(cosh\eta - cos\xi)^2}+\frac{2sinh^2\eta}{(cosh\eta - cos\xi)^3},\\
\frac{\partial}{\partial\xi}\frac{1}{cosh\eta - cos\xi}=-\frac{sin\xi}{(cosh\eta - cos\xi)^2},\\
\frac{\partial^2}{\partial\xi^2}\frac{1}{cosh\eta - cos\xi}=-\frac{cos\xi}{(cosh\eta - cos\xi)^2}+\frac{2sin^2\xi}{(cosh\eta - cos\xi)^3},
\end{gathered} \quad (12)$$

we obtain

$$\left(\frac{\partial^2}{\partial\eta^2}+\frac{\partial^2}{\partial\xi^2}\right)\left(\frac{1}{cosh\eta - cos\xi}\right)=\frac{cosh\eta + cos\xi}{(cosh\eta - cos\xi)^2}. \quad (13)$$

Then from Equations (11) to (13), we obtain

$$\begin{aligned}
&\left(\frac{\partial^2}{\partial\eta^2}+\frac{\partial^2}{\partial\xi^2}\right)\left(\frac{f(\xi)}{cosh\eta - cos\xi}\right)\\
&=\frac{(cosh\eta + cos\xi)f(\xi)-2sin\xi f'(\xi)+(cosh\eta - cos\xi)f''(\xi)}{(cosh\eta - cos\xi)^2}.
\end{aligned} \quad (14)$$

If we set $f''(\xi)=-f(\xi)$, Eq. (14) becomes

$$\left(\frac{\partial^2}{\partial\eta^2}+\frac{\partial^2}{\partial\xi^2}\right)\left(\frac{f(\xi)}{cosh\eta - cos\xi}\right)=\frac{2cos\xi f(\xi)-2sin\xi f'(\xi)}{(cosh\eta - cos\xi)^2}. \quad (15)$$

If we assume that $f(\xi)=Acos\xi + Bsin\xi$, then

$$\begin{gathered}
2cos\xi f(\xi)-2sin\xi f'(\xi)\\
=2Acos^2\xi+2Bsin\xi cos\xi+2Asin\xi^2-2Bsin\xi cos\xi=2A
\end{gathered} \quad (16)$$

Substituting Eqs. (15) and (16) into Eq. (10), we obtain

$$\frac{(cosh\eta - cos\xi)^2}{a^2}\left(\frac{\partial^2 u_s}{\partial \eta^2}+\frac{\partial^2 u_s}{\partial \xi^2}\right)-\frac{k}{\mu}=\frac{2A}{a^2}-\frac{k}{\mu}=0 \quad (17)$$

or

$$A=\frac{a^2k}{2\mu}. \quad (18)$$

The steady state solution $u_s$ of Eq. (1) satisfies no-slip boundary condition:

$$u_s(\xi_*,\eta)=0, u_s(2\pi-\xi_*)=0. \quad (19)$$

Therefore, we have

$$B = -Acot\xi_* = -\frac{a^2k}{\mu}cot\xi_*. \quad (20)$$

Then the steady flow is given by

$$u_s(\xi,\eta) = -\frac{ka^2}{2\mu}\frac{sin(\xi-\xi_{*j})}{sin(\xi_{*j})(cosh\eta - cos\xi)}, \quad (21)$$

with $\xi_{*j} = \xi_*$ for $\xi_* \leq \xi \leq \pi$ and $\xi_{*j} = 2\pi - \xi_*$ for $\pi \leq \xi \leq 2\pi - \xi_*$. The steady flow reduces to that of circular cylindrical tube $u_{circular}$ of equivalent diameter when $\xi_* = \pi/2$, $\xi = \pi$ and $\eta = 0$, given by [7]

$$u_{circular}(x=0, y=0) = -\frac{ka^2}{4\mu}. \quad (22)$$

The Poiseuille volumetric flow is then given by

$$Q_s = \frac{k}{4\mu}[a^4cot\xi_*\left(\frac{4}{3}+\frac{4}{3}cot^2\xi_* - 4csc^2\xi_*\right) + a^4csc^4\xi_*\left(-\xi_* - \frac{2}{3}sin2\xi_* - \frac{1}{16}sin4\xi_*\right)]. \quad (23)$$

The volumetric flow becomes that of the circular cylindrical tube of equivalent diameter when $\xi_* = \pi/2$, yielding [7]

$$Q_s = Q_{circular} = -\frac{k\pi a^4}{8\mu}. \quad (24)$$

The steady state shear stress $\tau_{\xi s}(\xi,\eta)$ by the fluid on the wall in $\xi$ axis which is a minor axis is given by

$$\begin{aligned}\tau_{\xi s} &= -\mu\frac{cosh\eta - cos\xi}{a}\frac{\partial u_s(\xi,\eta)}{\partial\xi} \\ &= \frac{ka}{2sin\xi_{*j}}\left[cos(\xi-\xi_{*j}) - \frac{sin(\xi-\xi_{*j})sin\xi}{cosh\eta - cos\xi}\right],\end{aligned} \quad (25)$$

and the steady state wall shear stress (WSS) is given by

$$\tau_{ws} = \tau_s\,|_{\xi_{*j}} = \frac{ka}{2sin\xi_{*j}} \quad (26)$$

which can be compared with that of the circular cylindrical tube with the equivalent diameter [7]

$$\tau_{circular} = \frac{ka}{2}. \quad (27)$$

One can assume that the circular cylindrical tube with the equivalent diameter $a$ corresponds to the case of a normal blood vessel of cylindrical geometry.

On the other hand, the steady state shear stress $\tau_{\eta s}(\xi,\eta)$ by the fluid on the wall in $\eta$ axis which is a major axis is given by

$$\tau_{\eta s} = -\mu\frac{cosh\eta - cos\xi}{a}\frac{\partial u_s(\xi,\eta)}{\partial\eta} = -\frac{kasin(\xi-\xi_*)}{2sin\xi_{*j}}\left[\frac{\sinh(\eta)}{cosh\eta - cos\xi}\right], \quad (28)$$

which becomes zero when $\xi = \xi_*$.

In order to find the oscillatory velocity profile $U_\phi$, we assume the trial solution of the form

$$U_\phi(\eta,\xi) = u_s(\eta,\xi)g(\eta,\xi). \quad (29)$$

Then by substituting Eq. (26) into Eq. (9), we obtain

$$\begin{aligned}\frac{i\omega\rho}{\mu}u_s g = \frac{(cosh\eta - cos\xi)^2}{a^2}\left[\left(\frac{\partial^2 u_s}{\partial\eta^2}+\frac{\partial^2 u_s}{\partial\xi^2}\right)g + 2\frac{\partial u_s}{\partial\xi}\frac{\partial g}{\partial\xi} + 2\frac{\partial u_s}{\partial\eta}\frac{\partial g}{\partial\eta\ +}\partial\eta\right.\\ \left. + \left(\frac{\partial^2 g}{\partial\eta^2}+\frac{\partial^2 g}{\partial\xi^2}\right)u_s\right] - \frac{k}{\mu}\\ \simeq \frac{k}{\mu}(g-1),\end{aligned} \tag{30}$$

where we assumed that $g$ is a slowing varying function and its derivatives can be ignored to the first order approximation. Then the oscillatory component of the velocity profile can be expressed as

$$U_\phi \simeq \frac{ku_s}{k - i\omega\rho u_s} \simeq \frac{ku_s}{k - i\omega\rho U_s}, \tag{31}$$

where $U_s$ is the peak value of the steady state velocity. Then the oscillatory volume flow can be approximated as

$$Q_\phi \simeq Q_s\left(\frac{k}{k - i\omega\rho U_s}e^{i\omega t}\right). \tag{32}$$

Then the real part of the oscillatory velocity profile and volume flow are given by

$$\begin{aligned}\Re u_\phi(\eta,\xi,t) = u_s(\eta,\xi)\frac{cos\omega t + \frac{\lambda}{4}sin\omega t}{1+(\frac{\lambda}{4})^2},\\ \Re Q_{phi} = Q_s\left(\frac{cos\omega t + \frac{\lambda}{4}sin\omega t}{1+(\frac{\lambda}{4})^2}\right),\end{aligned} \tag{33}$$

respectively. Here non-dimensional frequency parameter $\lambda$ is defined by

$$\lambda = \frac{\rho\omega a^2}{\mu} \tag{34}$$

which can be regarded as a Raynold number based on the frequency of oscillation $\omega$.

Likewise, the real part of the oscillatory WSS is given by

$$\Re\tau_{\phi w} = \tau_w\left[\frac{cos\omega t + \frac{\lambda}{4}sin\omega t}{1+(\frac{\lambda}{4})^2}\right]. \tag{35}$$

In pulsatile flow, the frequency parameter $\lambda$ plays a crucial role in governing fluid motion by influencing the interaction between viscous and inertial forces. In Womersley's foundational work, this parameter was approximated to $\lambda = 10$ based on physiological conditions in humans to accurately model blood flow dynamics. In our analysis, we adopt this value to illustrate key results and further vary it to $\lambda = 1$ examine how changes in frequency affect flow characteristics, such as velocity profiles, shear stress distribution, and wave propagation within the system.

In the following, we compare the flow properties of blood flow through bipolar cross section (BCS) and the normal cylindrical cross section (NCS) with equivalent diameters to understand the impact of these differences on hemodynamics.

Figure 2 illustrates the steady flow through the bipolar-shaped orifice, depicting the wall boundaries at positions (a) $\xi_* = 2\pi/3$, (b) $\xi_* = 3\pi/4$, (c) $\xi_* = 4\pi/5$, and (d) $\xi_* = 5\pi/6$, respectively. The velocity profile is normalized to the peak velocity observed in the NCS with an equivalent diameter. This normalization allows for a direct comparison between the flow characteristics of the BCS and NCS, ensuring that differences in their respective velocity profiles are highlighted relative to a common reference point. The velocity profiles for the BCS show a reasonable agreement with those obtained in previous studies using coherent multi-scale simulations [6]. These profiles consistently demonstrate the presence of a jet-like flow structure within the fluid, a feature that is notably absent in the NCS scenarios. This jet formation is indicative of the distinct hemodynamic patterns associated with BCS, underscoring the significant impact of valve morphology on flow dynamics.

Figure 3 presents comparative velocity profiles at the aorta entrance for (a) BCS, (b) NCS, and (c) a combined profile showing BCS (red) and NCS (blue) at the center. The analysis reveals that at the center of the entrance, the velocity for the BCS is significantly higher compared to the case of NCS. However, the BCS velocity decreases more rapidly than that of the NCS as it moves towards the vessel wall. This rapid decrease in velocity for the BCS creates a steeper velocity gradient in the vertical direction towards the vessel wall. Consequently, this results in higher wall shear stress in the case of BCS. The increased wall shear stress can have significant implications for vascular health, potentially influencing the development of aortic diseases and complications associated with BAV. Our results demonstrate a reasonably good agreement with Figure 6 of Reference 6. In this reference, the computation time was on the order of minutes for each cell, whereas our analytical model-based computation achieves similar results in just a few seconds. This significant reduction in computation time highlights the efficiency and effectiveness of our approach, providing rapid and reliable analysis that can be advantageous for both research and clinical applications.

Figure 4 presents the normalized velocity profiles ($u_\phi/U_s$) at both high and low frequencies, illustrating key differences in flow behavior across the oscillatory cycle for the case of $\xi_* = 5\pi/6$. The results are displayed for the first half of the cycle, spanning from ($\omega t = 0$ to $\omega t = \pi$) as the second half ($\omega t = \pi$ to $\omega t = 2\pi$) is simply a mirror image, reflected across the zero normalized velocity line due to the periodic nature of pulsatile flow. At low frequencies, the velocity distribution maintains a relatively simple structure, with the peak velocity consistently aligned along the longitudinal axis of the tube throughout the oscillatory phase. The maximum velocity occurs near $\omega t = 0$ and $\omega t = \pi$, corresponding to the points of highest forward and backward flow in the cycle. The overall velocity profile remains smooth, with gradual acceleration and deceleration phases. In contrast, at high frequencies, the influence of oscillation frequency on velocity distribution becomes more pronounced. Instead of obtaining its peak along the longitudinal axis at $\omega t = 0$, the velocity shifts toward maximum during the initial stages of the cycle ($\omega t = 0$ to $\omega t = \pi/2$). This shift is indicative of increased flow asymmetry and complex secondary flow patterns that arise due to higher oscillatory inertia. Following this phase, the velocity declines sharply as the cycle progresses toward $\omega t = \pi$, demonstrating a more abrupt transition compared to the low-frequency case. These frequency-dependent variations in velocity distribution highlight the critical role of oscillation frequency in shaping flow characteristics, with potential implications for hemodynamic studies and cardiovascular fluid dynamics.

The results of normalized volume flow rate ($Q_\phi/Q_s$) is figure 5 for wide range of frequency parameter $\lambda$. At low frequencies, the maximum volume flow occurs near $\omega t = 0$ and $\omega t = \pi$, and at high frequencies, the volume flow shifts toward maximum during the initial stages of the cycle ($\omega t = 0$ to $\omega t = \pi/2$) and becomes nearly constant at very high frequency ($\lambda = 100$).

Figure 6 illustrates the normalized shear stress distribution across the bipolar-shaped orifice, highlighting the wall boundaries at positions (a) $\xi_* = 2\pi/3$, (b) $\xi_* = 3\pi/4$, (c) $\xi_* = 4\pi/5$, and (d) $\xi_* = 5\pi/6$. As derived from equation (7) and discussed in the context of Figure 3, the wall shear stress (WSS) reaches its maximum at the boundary of the bicuspid aortic valve. The analysis reveals that as the shape of the bicuspid valve becomes narrower, the WSS increases significantly. This indicates that the geometry of the bicuspid valve has a critical impact on the shear stress experienced at the vessel wall, with narrower valve shapes leading to higher shear stress. This finding is essential for understanding the hemodynamic stresses associated with bicuspid aortic valves and their potential implications for vascular health.

In figure 7, we plot WSS of BCS normalized by the WSS of NCS. The normalized WSS is inversely proportion to $sin(\xi_*)$ which is rapidly increasing as the orifice of the aortic valve becomes more asymmetrical.

In Figure 7, we plot the wall shear stress (WSS) of the BCS normalized by the WSS of the NCS. The normalized WSS is inversely proportional to $sin(\xi_*)$ of the aortic valve orifice, rapidly increasing as the orifice becomes more asymmetrical. This demonstrates that as the aortic valve deviates from a symmetric shape, the WSS increases significantly, which can have important implications for the structural integrity and function of the valve.

Under steady flow conditions maximum shear stress occurs at the minor axis ($\xi$ axis) while the minimum shear stress occurs on the major axis ($\eta$ axis). It follows that for pulsatile flow, at low frequency, maximum stress occurs on the minor axis at every point in the phase of oscillation of the fluid while minimum occurs on the major axis as can be seen in Figure 8. As the frequency increases, more complex behavior is observed. The result is a delay in the reaction of fluid to changing near the surface the bipolar cross section along $\xi$ coordinate.

Under steady flow conditions, the distribution of shear stress within the bipolar cylindrical cross-section follows a predictable pattern: the maximum shear stress is concentrated along the minor axis ($\xi$ axis), while the minimum shear stress occurs along the major axis ($\eta$ axis). This distribution arises due to the geometric asymmetry of the cross-section, influencing how shear forces interact with the vessel walls. For pulsatile flow at low frequencies, this trend remains consistent throughout the oscillatory cycle. At every phase of fluid motion, the peak shear stress continues to occur along the minor axis, while the major axis experiences the lowest shear stress. This behavior, illustrated in Figure 8, suggests that at low oscillatory frequencies, the fluid has sufficient time to adjust to the changing pressure gradients, maintaining a relatively stable shear stress distribution that mirrors steady-flow conditions. However, as the oscillation frequency increases, the shear stress distribution exhibits more complex behavior. Higher frequencies introduce greater inertia and unsteady effects, leading to delays in the fluid's response to oscillatory pressure changes. Near the vessel wall, particularly along the bipolar cross-section in the $\xi$-coordinate direction, this delay manifests as a phase lag in the shear stress response. Essentially, the fluid near the boundary does not immediately react to the imposed oscillations, creating regions where shear stress is temporarily out of sync with the driving forces of the flow. This phase lag becomes increasingly pronounced at higher frequencies, highlighting the interplay between inertial effects and wall-bounded flow dynamics in pulsatile hemodynamics. Such frequency-dependent shear stress variations are of particular interest in cardiovascular biomechanics, where oscillatory flow patterns influence endothelial cell function, arterial remodeling, and disease progression, such as in the development of atherosclerotic plaques. Understanding these effects within non-circular geometries can provide deeper insights into the hemodynamic forces acting on blood vessels under physiological and pathological conditions.

## III. Summary

Pulsatile flow through compressed or defective blood vessels plays a crucial role in hemodynamics, particularly in cardiovascular studies. This research examines flow dynamics within a tube with a bipolar cross-section, representing the geometry of bicuspid aortic valves (BAV), aortic bifurcations, and the aortic arch—regions where non-uniform vessel shapes significantly affect hemodynamic patterns.

In this study, we derive an analytical solution to the governing equations for both Poiseuille and pulsatile flow in a bipolar cross-section. The analysis focuses on velocity distribution, flow rate, and wall shear stress (WSS). At low frequencies, the velocity profile remains smooth, with gradual acceleration and deceleration phases, while at higher frequencies, oscillatory effects become more pronounced. The maximum volume flow occurs near $\omega t = 0$ and $\omega t = \pi$ at low frequencies, but at higher frequencies, it shifts toward an early peak in the cycle ($\omega t = 0$ to $\omega t = \pi/2$) before stabilizing at very high frequencies. Shear stress behavior also varies with frequency. At low oscillatory frequencies, the fluid adapts smoothly to changing pressure gradients, resulting in a shear stress distribution similar to steady flow. However, as frequency increases, inertial and unsteady effects introduce phase lags in the fluid's response, leading to more complex shear stress distributions. These findings provide deeper insight into how vessel geometry and pulsatile forces influence cardiovascular hemodynamics, with potential applications in understanding disease progression and improving diagnostic models.

**References**

1. Uchida, S. The pulsating viscous flow superimposed on the steady laminar motion of incompressible fluid in a circular pipe. Z. Angew. Math. Phys. 7, 403-422 (1956).
2. Womersley, J. R. Oscillatory motion of a viscous liquid in a thin-walled elastic tube-I: The linear approximation for long waves. Philos. Mag. 46, 199-221 (1955).
3. Khamrui, S. R. On the flow of a viscous liquid through a tube of elliptic section under the influence of a periodic gradient. Bull. Calcutta. Math. Soc. 49, 57-60 (1957).
4. Haslam, M. & Zamir, M. Pulsatile flow in tubes of elliptic cross sections. Ann. Biomed. Engin. 26, 780-787 (1998).
5. Fanucci, E., Orlacchio, A. & Pocek, M. The vascular geometry of human arterial bifurcations. Invest. Radiol. 23, 713-718 (1988).
6. Abhilash, H. N., Yanagita, Y., Pal, R., Zuber, M., Tamagawa, M., Prakashini, K., Ganesh, K. S., Padmakumar, R., Barboza, A. B. V. , Rao, V. R. K. & Abdul Khader, S. M. Effect of vascular geometry on hemodynamic changes in a carotid artery bifurcation using numerical simulation. Clinic. Nerol. Neurosurg. 237, 108153 (2024).
7. Rodriguez-Palomares, J. F., Dux-Santoy, L., Guala, A., Galian-Gay, L. & Evangelista, A. Mechanisms of aortic dilation in patients with bicuspid aortic valve. J. Am. College Cardio. 82, 448-464 (2023).
8. Michelena, H. I. Et al. International consensus statement on nomenclature and classification of the congenital bicuspid aortic valve and its aortopathy, for clinical, surgical, interventional and research purposes. European J. Cardio-Thorac. Surg. 60, 448-476 (2021).
9. Girdauskas, E., Borger, M. A. Sechnus, M.-A., Girdauskas, G. & Kuntze, T. Is aortopathy in bicuspid aortic valvue lease a congenital or a result of abnormal hemodynamics? Critical reappraisal of a one-sided argument. European J. Cardio-Thoracic Surgery 39, 809-814 (2011).
10. Conti, C. A., Cortie, A. D., Votta, E., Viscose, L. D., Bancone, C., De Santo, L. S. & Redaelli, A. Biomechanical implications of the congenital bicuspid aortic valve: A finite element study of aortic root function from in vivo data. J. Thorac. Card. Surg. 140, 890-896 (2010).
11. Robcsek, F., Thubrikar, M. J., Cook, J. W. & Fowler, B. The congenitally bicuspid aortic valve: How does it function? Why does it fail? Ann. Thorac. Surg. 77, 177-185 (2004).
12. Weinberg, E. J. & Moored, M. R. K. A multiscale computational comparison of the bicuspid and tricuspid aortic valves in relation to aortic stenosis. J. Biomechanics 41, 3482-3487 (2008).
13. Zamir M. Hemo-Dynamics. London, UK:Springer (2015).
14. Malsen, S. H. "Transverse velocities in fully developed flows", Q. Appl. Math. 16, 173-175 (1958)

15. Lee, Y. Y. & Ahn, D. Dispersive full-wave finite-difference time-domain analysis of the bipolar cylindrical cloak based on effective medium approach. J. Opt. Soc. Am. B 30, 140-148 (2013).
16. Moon, P. & Spencer, D. E. Field Theory for Engineers. Princeton, NJ:Van Nostrand, (1961).

**Acknowledgments**

This work is supported by 2024 Research grant from the University of Seoul.

**Data availability**

The data generated during the current study are available from the corresponding author on reasonable request.

**Competing interests**

The authors declare no competing interests.

**Figure legends**

**Figure 1.** Illustration of the bipolar coordinates.

**Figure 2**. Flow velocity profiles of the steady flow through the bipolar cross sections, depicting the wall boundaries at positions (a) $\xi_* = 2\pi/3$, (b) $\xi_* = 3\pi/4$, (c) $\xi_* = 4\pi/5$, and (d) $\xi_* = 5\pi/6$, respectively. The velocity profile is normalized to the peak velocity observed in the normal cylindrical cross section (NCS) with an equivalent diameter.

**Figure 3.** Comparative velocity profiles at the aorta entrance for (a) bipolar cylindrical cross section (BCS), (b) normal cylindrical cross section (NCS), and (c) a combined profile showing BCS (red) and NCS (blue) at the center.

**Figure 4.** Velocity profiles along the normalized major ($x/a$) and minor ($y/atan\xi_*$) axes of a bipolar cross section with the wall boundary at $\xi_* = 5\pi/6$, at different times within the oscillatory cycle. Results for the low frequency for normalized major axis are shown in (a), the high frequency for the normalized major axis in (b), the low frequency for normalized minor axis in (c) and the high frequency for the normalized minor axis are in (d).

**Figure 5**. Variation of the normalized volume flow rate with in the oscillatory cycle in a tube of bipolar cross section for a wide range of frequency parameter $\lambda$.

**Figure 6**. Normalized shear stress distribution across the bipolar cross section, highlighting the wall boundaries at positions (a) $\xi_* = 2\pi/3$, (b) $\xi_* = 3\pi/4$, (c) $\xi_* = 4\pi/5$, and (d) $\xi_* = 5\pi/6$. As derived from equation (26) and discussed in the context of Figure 6, the wall shear stress (WSS) reaches its maximum at the boundary of the bipolar cross section.

**Figure 7.** Plot of the wall shear stress (WSS) of the bipolar cylindrical cross section (BCS) normalized by the WSS of normal cylindrical cross section (NCS).

**Figure 8.** The variation of the shear stress with in the oscillatory cycle for (a) low frequency ($\lambda = 1$) and high frequency ($\lambda = 10$). At low frequencies, shear stress resembles steady flow, but higher frequencies introduce inertia and phase lags, creating more complex patterns.

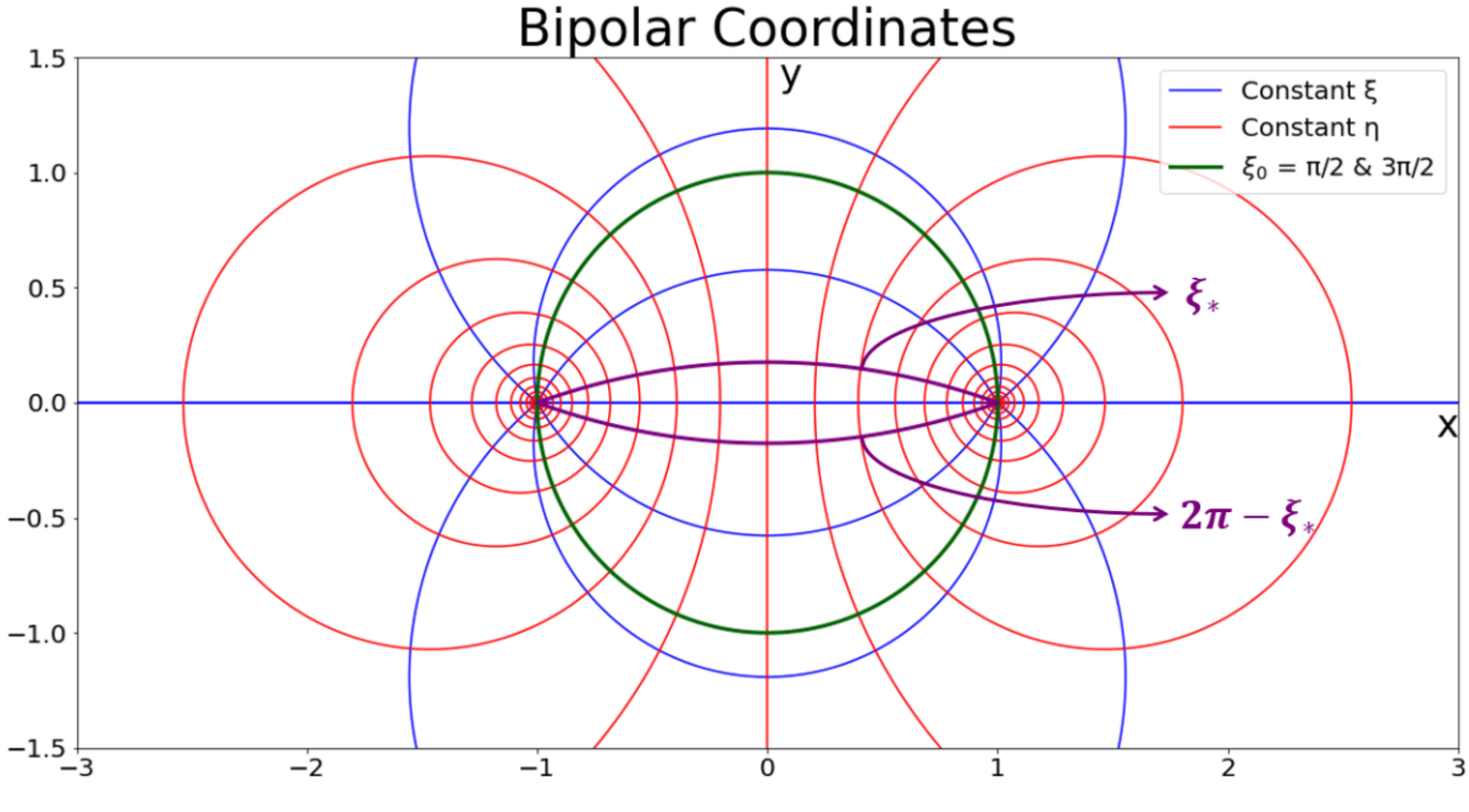

Fig. 1

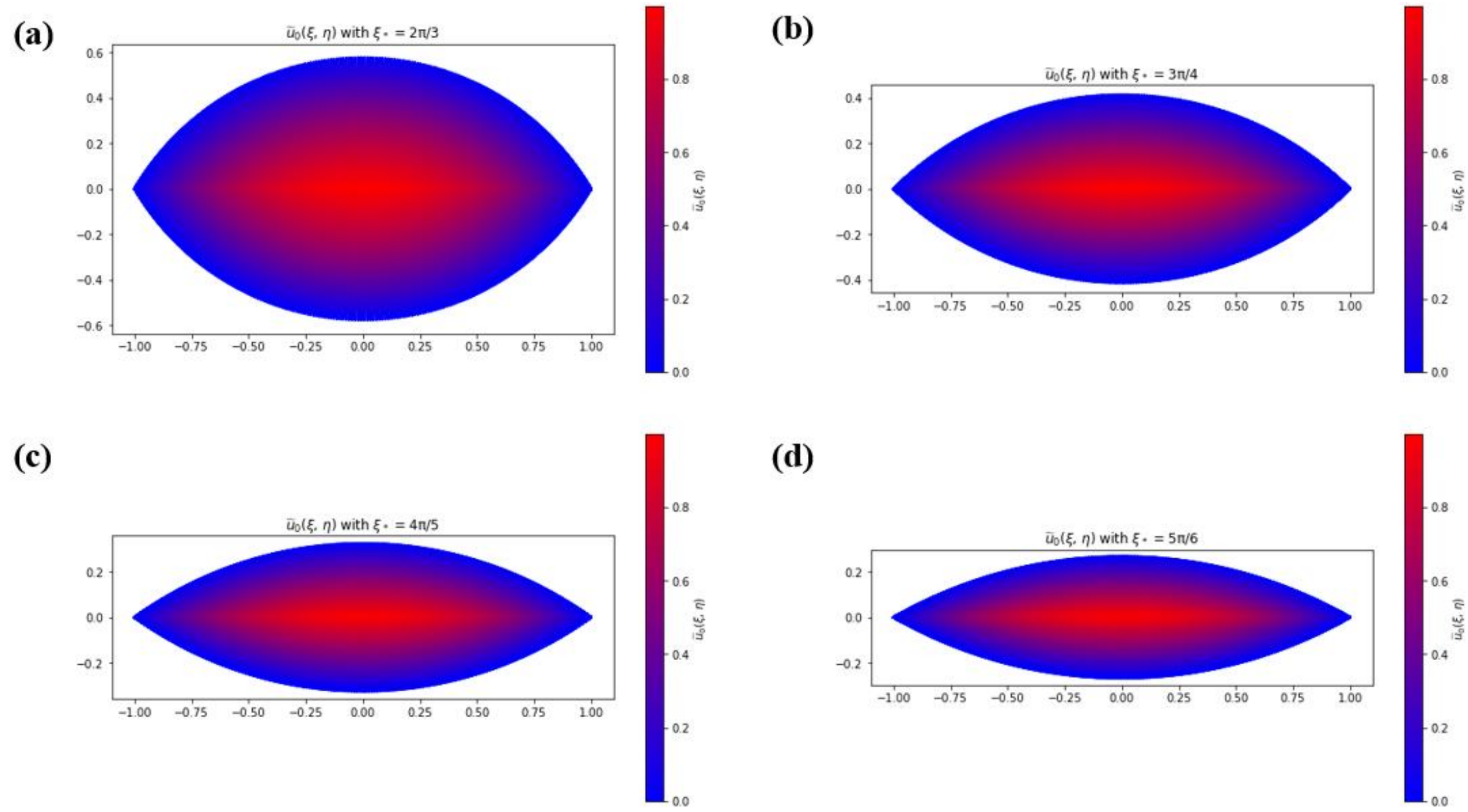

Fig. 2

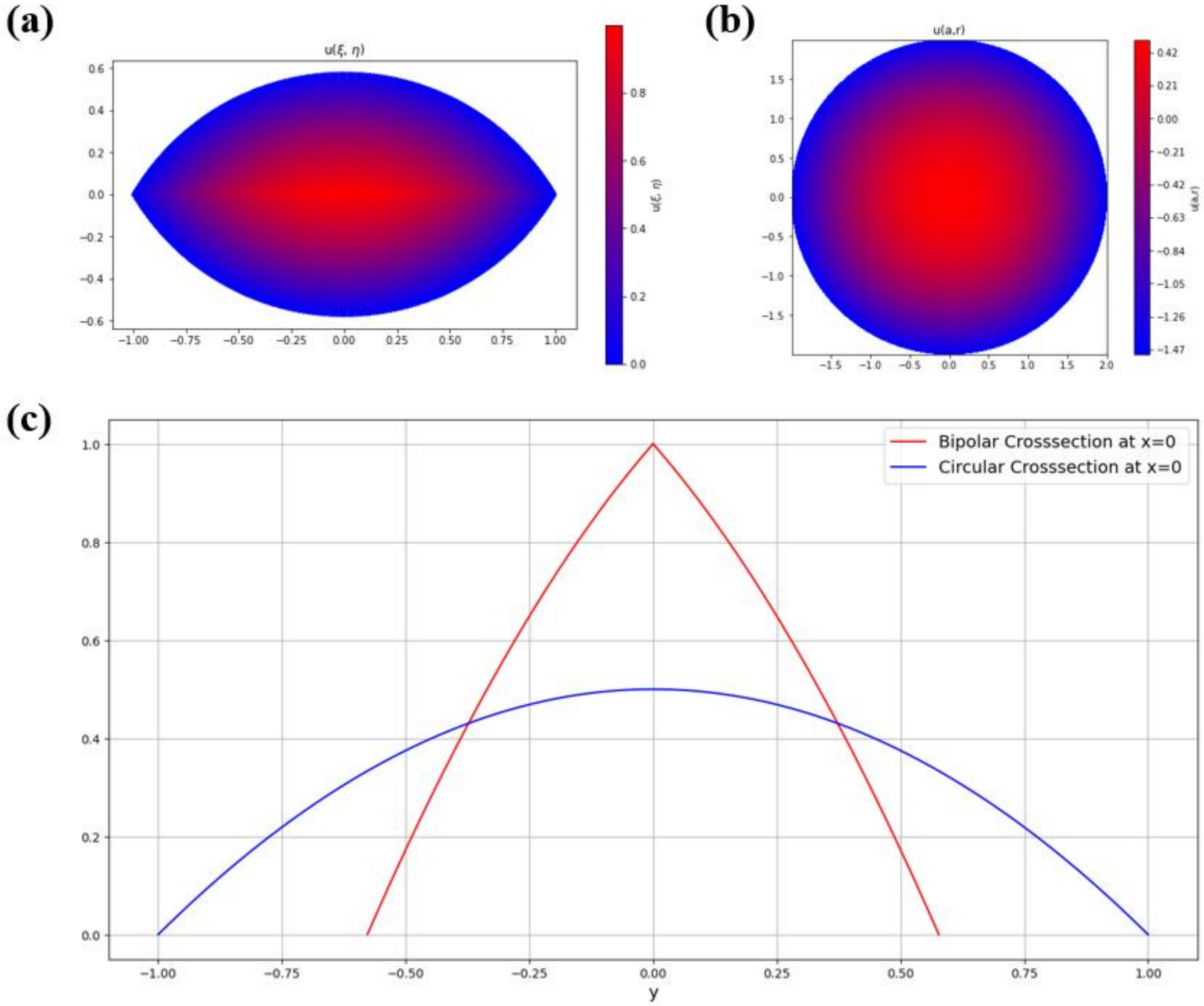

Fig. 3

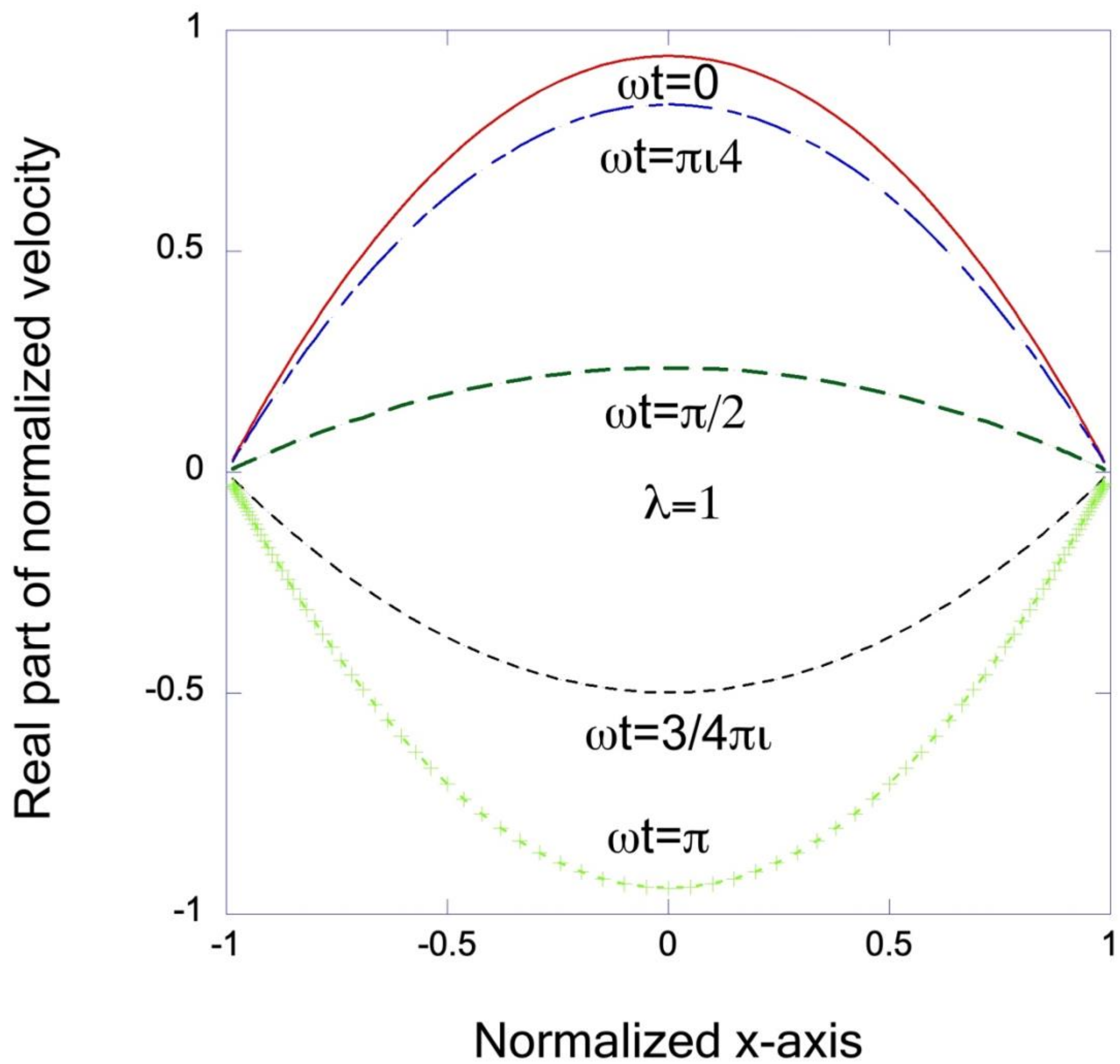

Fig. 4(a)

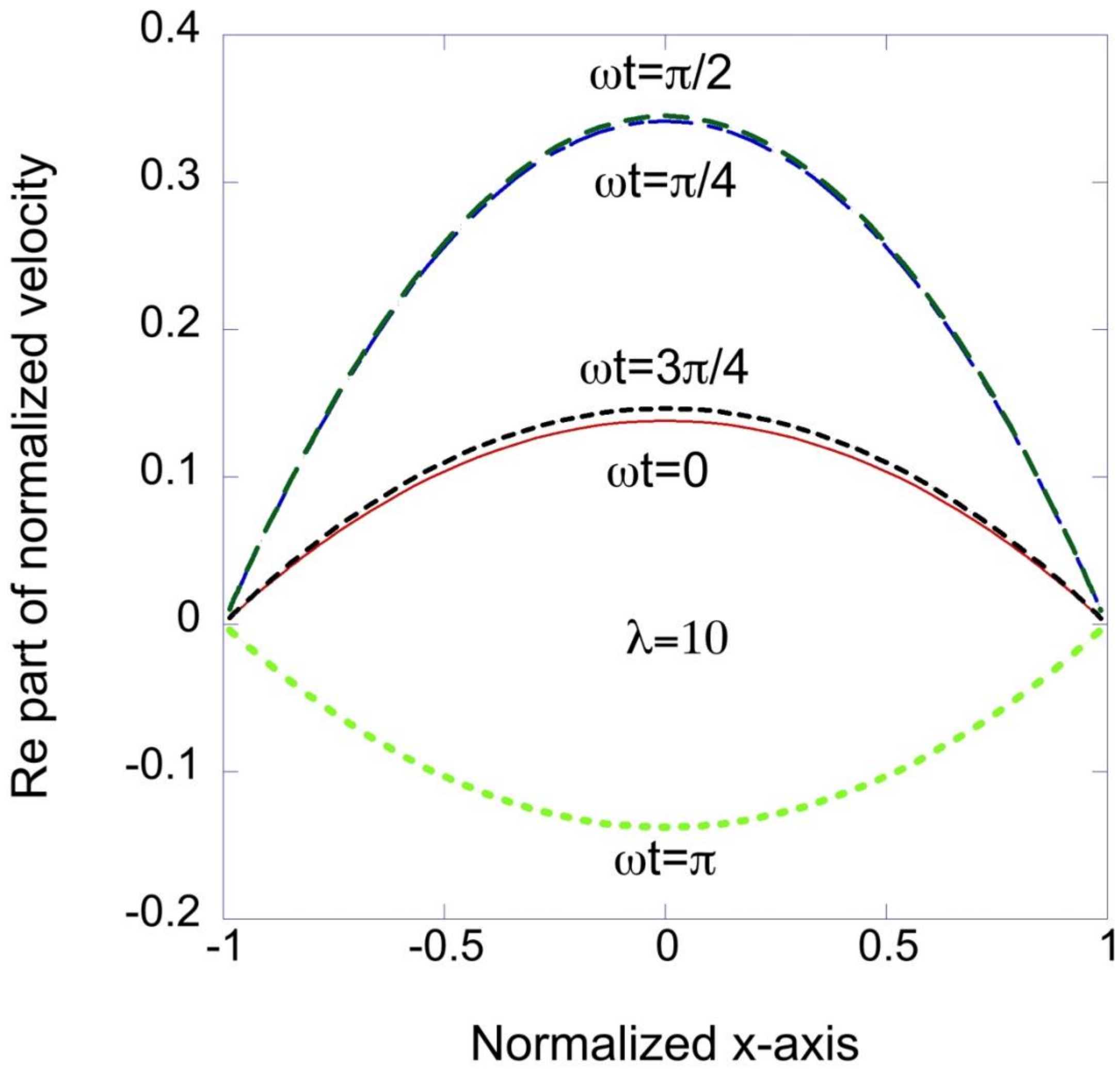

Fig. 4(b)

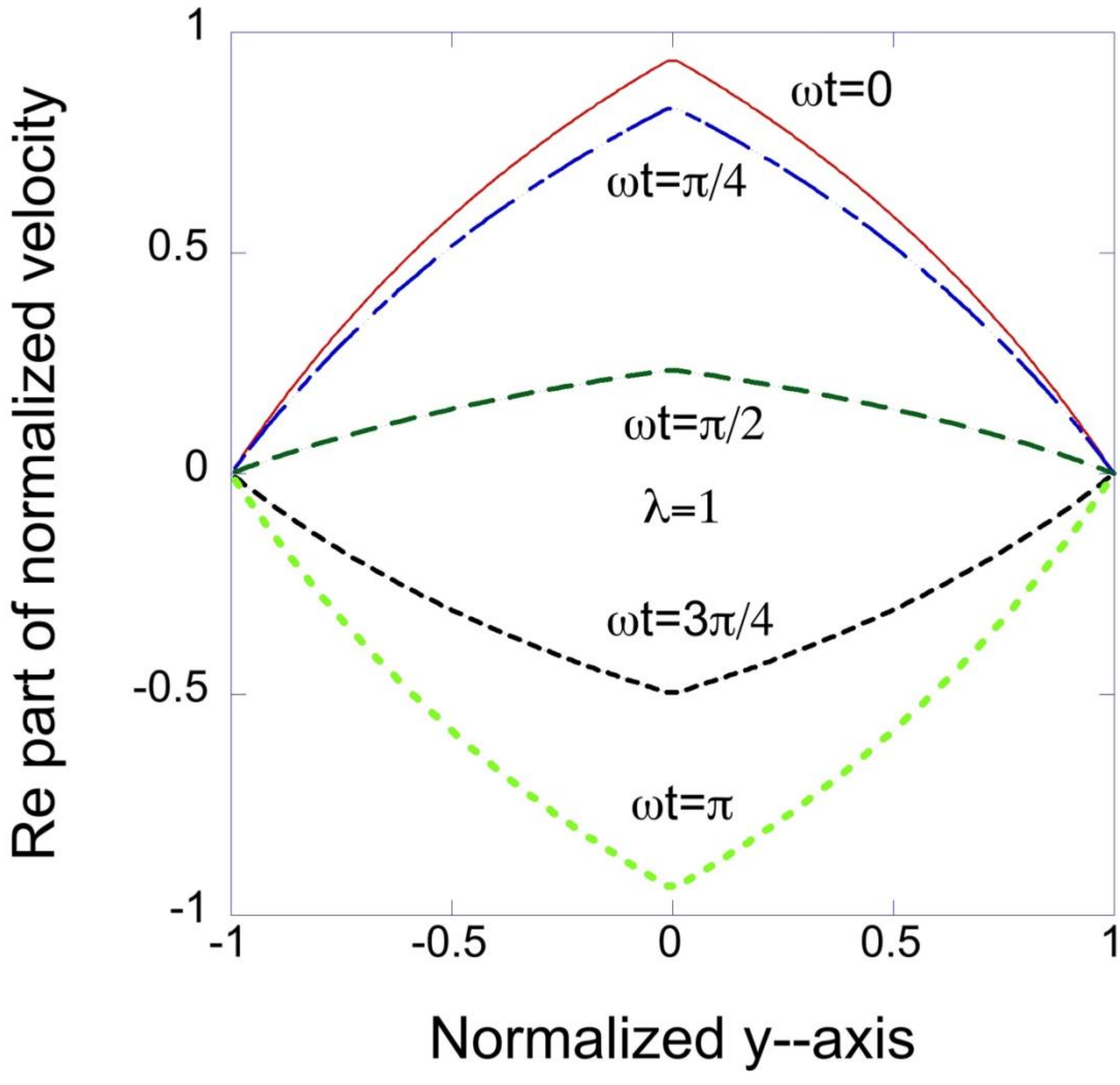

Fig. 4(c)

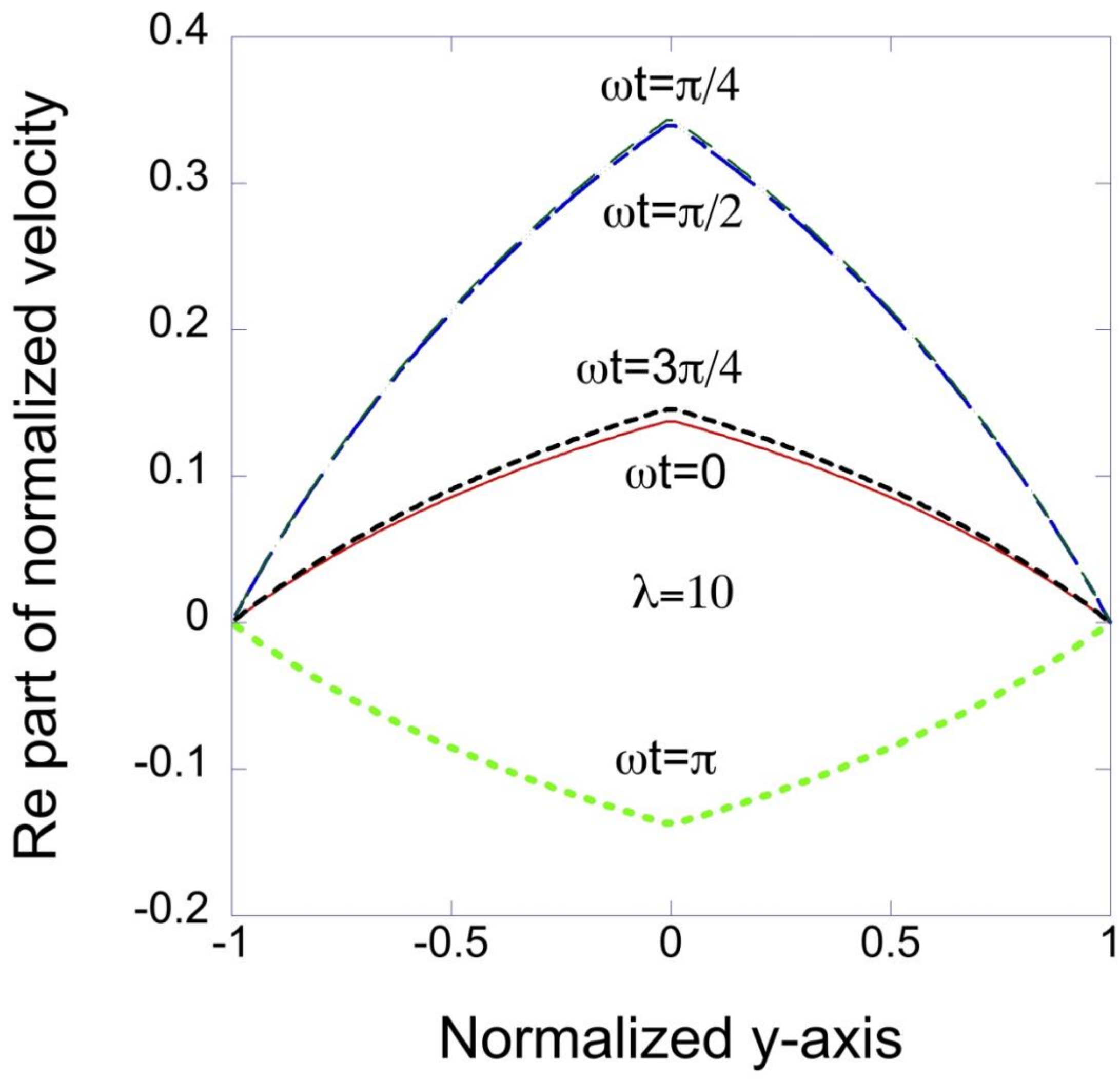

Fig. 4(d)

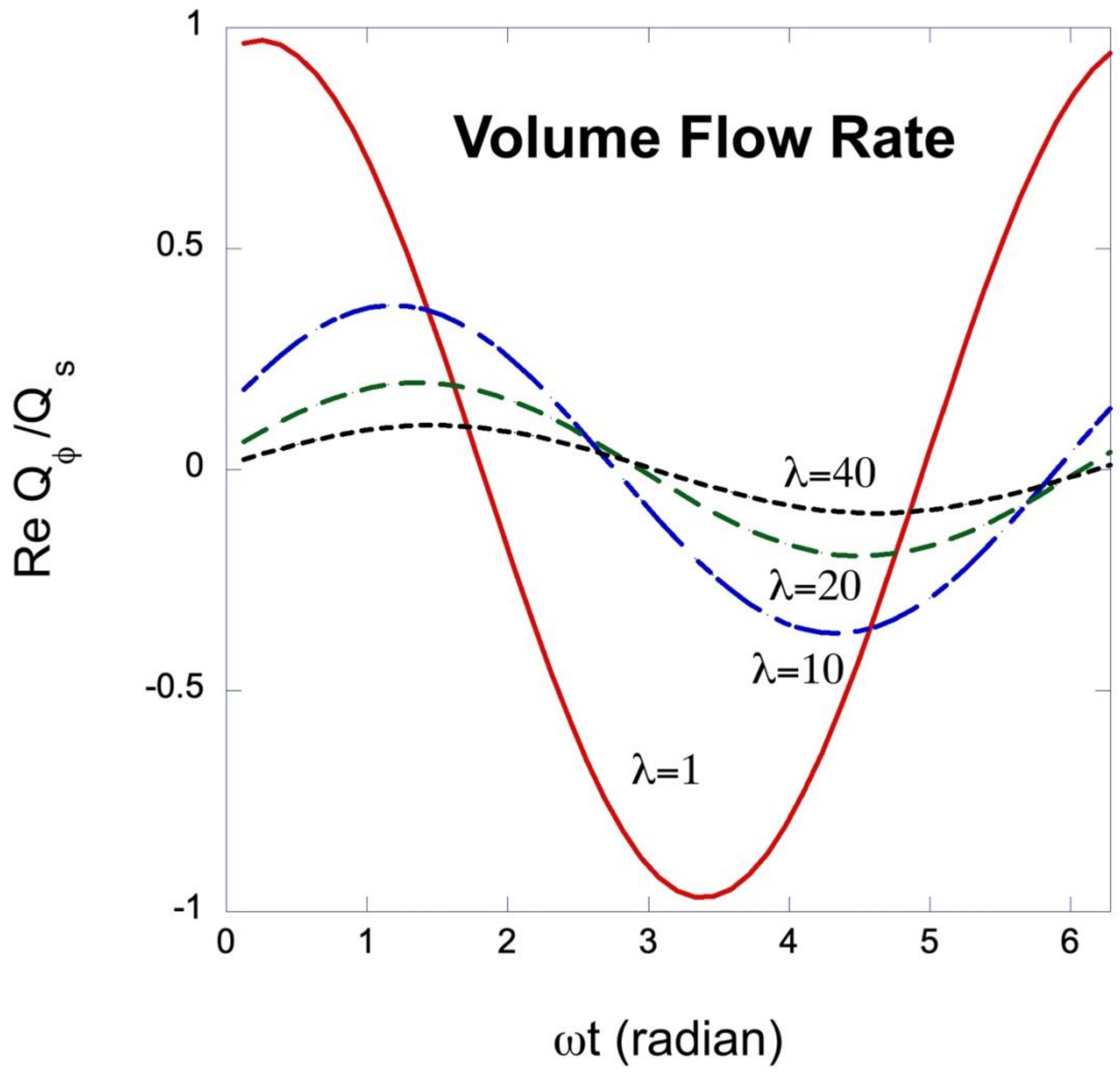

Fig. 5

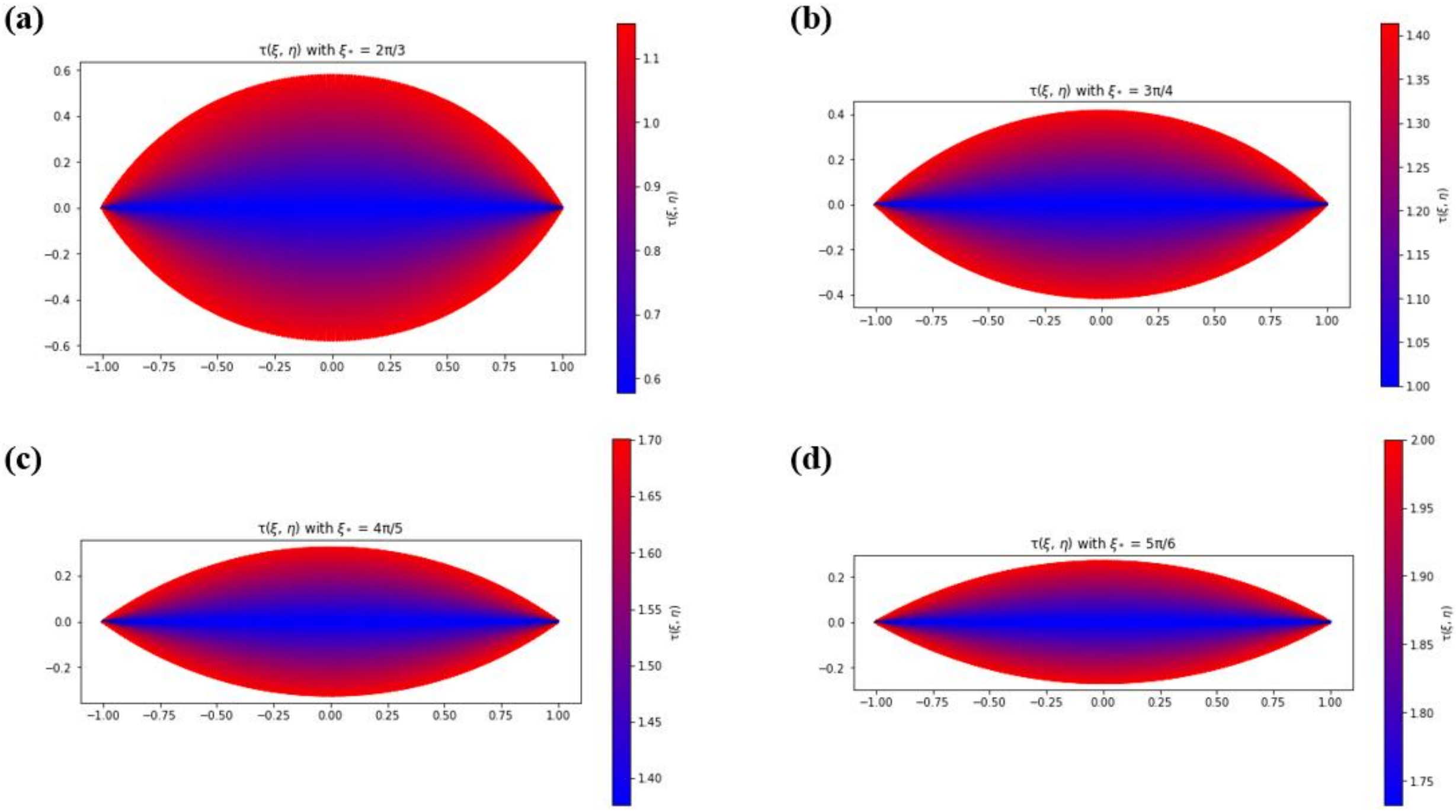

Fig. 7

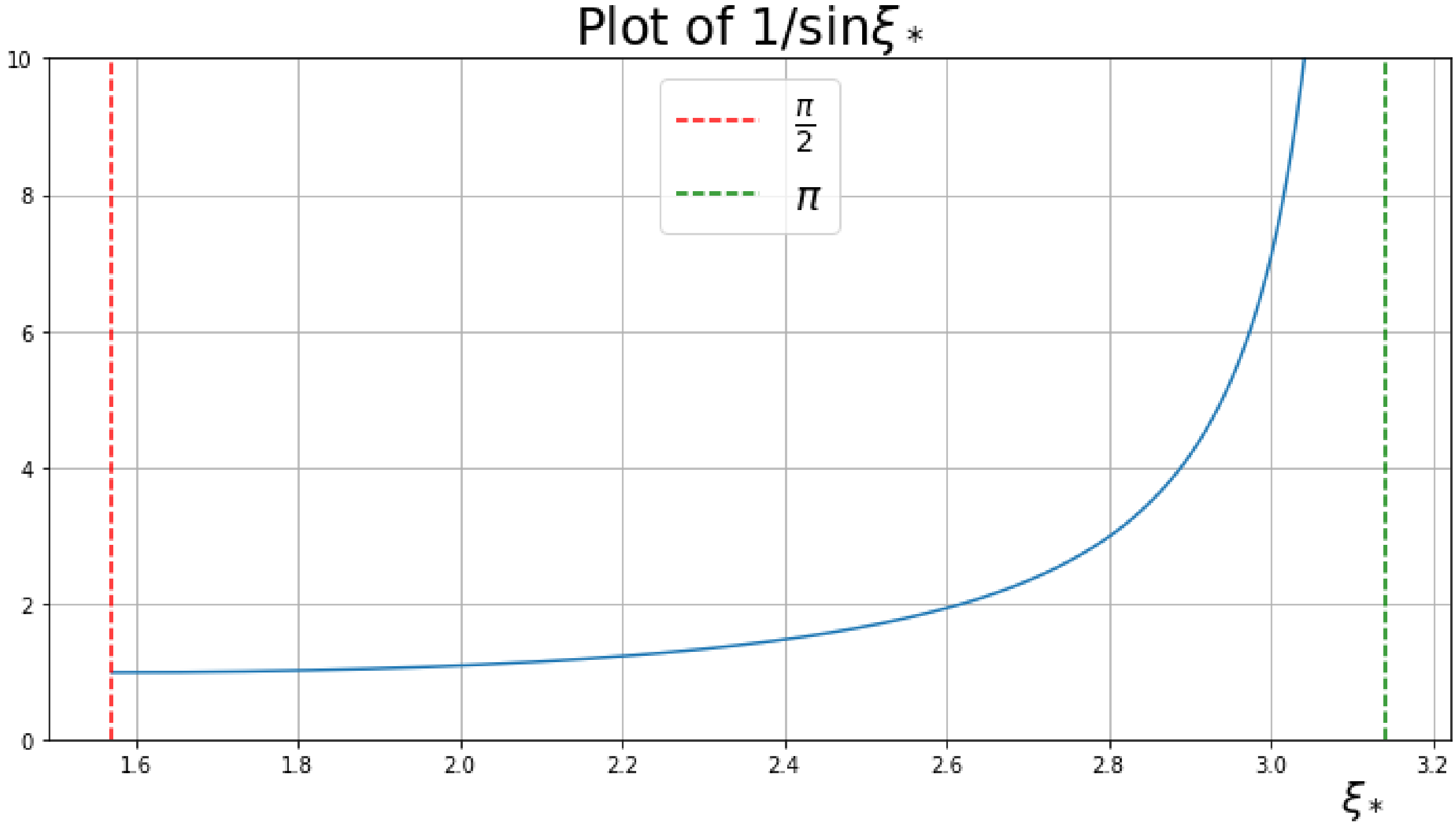

Fig. 7

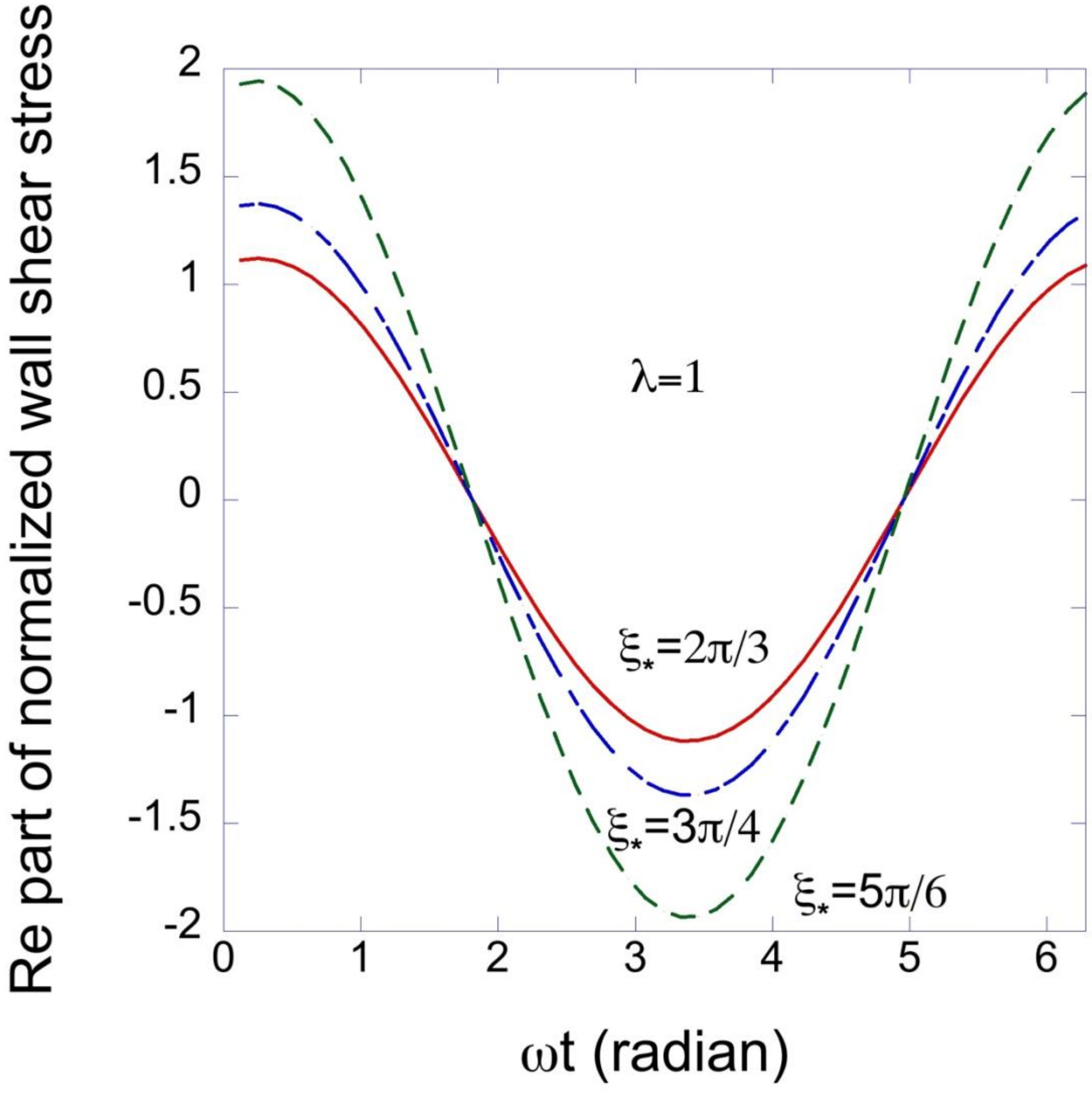

Fig. 8(a)

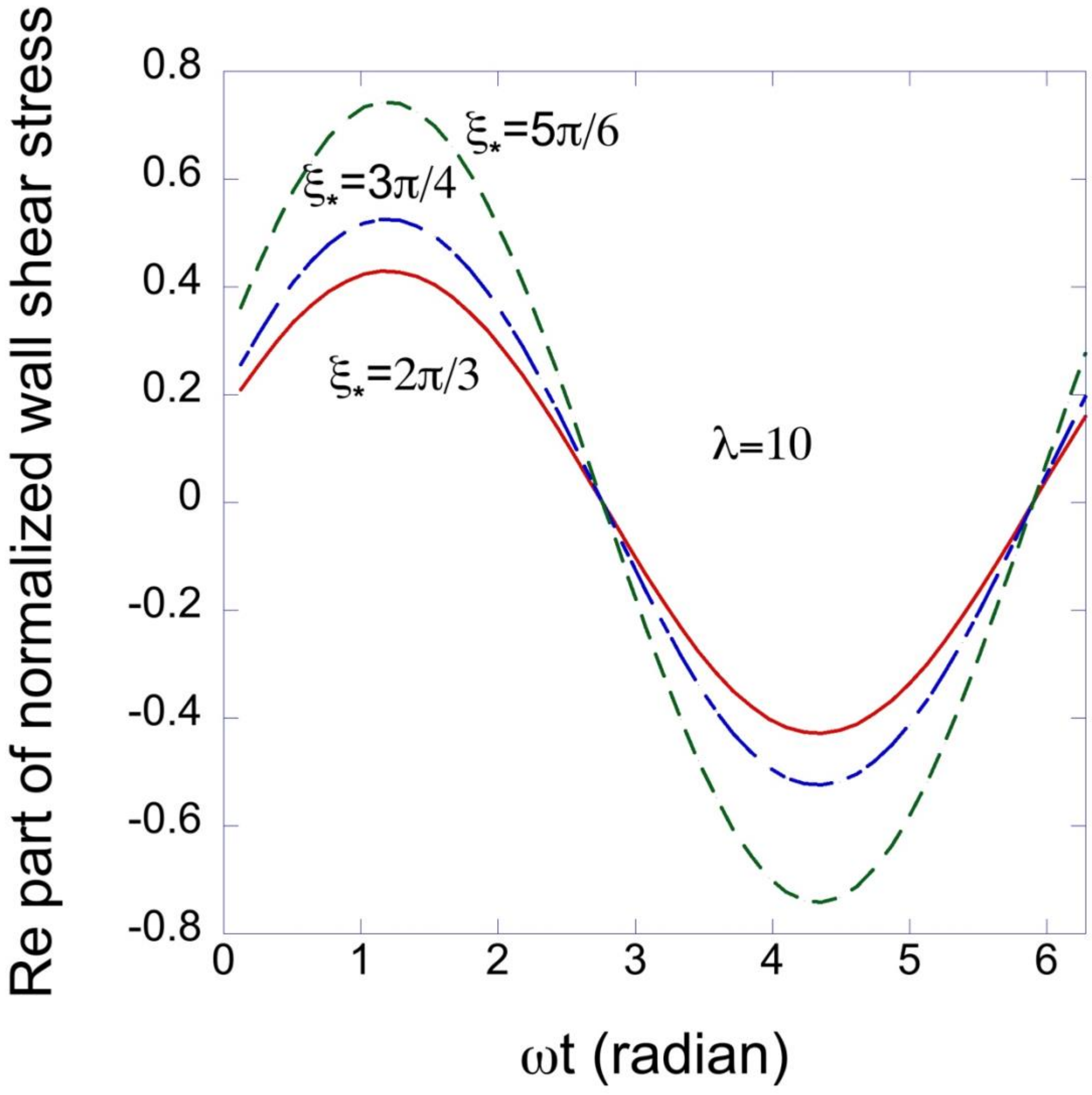

Fig. 8(b)